\documentstyle[aps,prb,multicol,epsf]{revtex}
\begin{document}
\draft
\preprint{}
\wideabs{
\title{Effect of magnetic frustration on single-hole spectral
function in the $t$-$t'$-$t''$-$J$ model}
\author{Y. Shibata, T. Tohyama and S. Maekawa}
\address{Institute for Materials Research, Tohoku University,
        Sendai 980-8577, Japan}
\date{\today}
\maketitle

\begin{abstract}
We examine the effect of the magnetic frustration $J'$ on
the single-hole spectral function in the $t$-$t'$-$t''$-$J$ model.
At zero temperature, the exact diagonalization (ED) and the
self-consistent Born approximation (SCBA) methods are used.
We find that the frustration suppresses the quasiparticle (QP)
weight at small momentum {\bf k}, whereas the QP peak at
{\bf k}=($\pi$/2,$\pi$/2) remains sharp.  We also show the
temperature dependence of the single-hole spectral function
by using the ED method.  It is found that the lineshapes at
($\pi$/2,0) and ($\pi$/2,$\pi$/2) show different temperature
dependence.  These findings are consistent with the angle-resolved
photoemission data on Sr$_2$CuO$_2$Cl$_2$, and indicate the
importance of the magnetic frustration on the electronic states
of the insulating cuprates.
\end{abstract}
\pacs{PACS numbers: 75.50.Ee, 78.20.Bh, 78.66.Nk, 79.60.-i}
}
\narrowtext

\section{Introduction}
\label{sec1}
Since the discovery of high temperature superconductivity,
the $t$-$J$ model has been extensively studied as a theoretical
model for the superconductors.  However, angle-resolved photoemission
spectroscopy (ARPES) measurements on Sr$_2$CuO$_2$Cl$_2$ by Wells
{\it et al.}\cite{Wells} have clearly shown that the $t$-$J$ model
does not explain the experimental data near ${\bf k}$=($\pi$,0):
the $t$-$J$ model predicts that the energies of the quasiparticle
(QP) at ($\pi$,0) and ($\pi$/2,$\pi$/2) are very similar, while the
experimental energy at ($\pi$,0) is lower than that at
($\pi$/2,$\pi$/2).  Recent theoretical
works\cite{Nazarenko,Kyung,Xiang,Belinicher,Lee,Eder,Leung1,Lema,Duffy}
have shown that this discrepancy can be resolved by introducing
hopping matrix elements to second and third nearest neighbors
($t'$ and $t''$) into the $t$-$J$ model.  The Hamiltonian, termed
the $t$-$t'$-$t''$-$J$ model, is given by
\begin{eqnarray}
H_{tt't''J}&=&H_{tt't''}+H_J\;,\\
\label{H}
H_{tt't''}&=&-t\sum\limits_{\left<i,j\right>_{1{\rm st}} \sigma }
    \tilde{c}_{i\sigma }^\dagger \tilde{c}_{j\sigma }
-t'\sum\limits_{\left<i,j\right>_{2{\rm nd}} \sigma }
    \tilde{c}_{i\sigma }^\dagger \tilde{c}_{j\sigma } \nonumber \\
&& {} -t''\sum\limits_{\left<i,j\right>_{3{\rm rd}} \sigma }
    \tilde{c}_{i\sigma }^\dagger \tilde{c}_{j\sigma }+{\rm H.c.}\;,\\
\label{Ht3}
H_J&=&J\sum\limits_{\left<i,j\right>_{1{\rm st}}} {{\bf S}_i}\cdot {\bf S}_j\;,
\label{HJ}
\end{eqnarray}
where $\tilde{c}_{i\sigma}=c_{i\sigma}(1-n_{i-\sigma})$ is the
annihilation operator of an electron with spin $\sigma$ at site $i$
with the constraint of no double occupancy, ${\bf S}_i$ is the spin
operator and the summations $\left< i,j \right>_{1{\rm st}}$,
$\left< i,j \right>_{2{\rm nd}}$ and $\left< i,j \right>_{3{\rm rd}}$
run over first, second and third nearest-neighbor pairs, respectively.

Very recently, it has been argued\cite{Kim} that not only the QP
dispersion but also the lineshape of the ARPES data contains valuable
information about the electronic states.  It was shown that the
$t$-$t'$-$t''$-$J$ model describes the observed broad peak at ($\pi$,0).
A weakening of the antiferromagnetic (AF) spin correlation induced by
the charge-carrier motion due to the longe-range hoppings is found to
be responsible for the broad lineshape.

The fact that the lineshape of ARPES spectra is sensitive to the AF
correlation in the spin background suggests that, if magnetic
frustration in the spin system is strong, the lineshape of the spectra
should be also affected significantly.  A dominant interaction which
causes the magnetic frustration is considered to be the next-nearest
neighbor exchange interaction $J'$.  The Hamiltonian is given by
\begin{equation}
H_{J'}=J'\sum\limits_{\left<i,j\right>_{2{\rm nd}}}
{{\bf S}_i}\cdot {\bf S}_j\;.
\label{HJ'}
\end{equation}
The two-dimensional (2D) spin-${1\over 2}$ Heisenberg Hamiltonian
including both $J$ and $J'$, i.e. $H_J$ + H$_{J'}$, has been
extensively studied, motivated by the suggestion\cite{Anderson} that
the physics of the high-$T_c$ phenomena is closely related to the
presence of non-N\'eel states in insulating 2D cuprates.  The phase
diagram of the $J$-$J'$ model obtained as a function of $J'$/$J$ is
as follows:\cite{Mila,Miyazawa}  the N\'eel state is the ground state
for 0$\leq$$J'$/$J$$\alt$0.6, while a collinear state becomes stable
in the ground state for 0.6$\alt$$J'$/$J$$<$1.  Since the realistic
value of $J'$/$J$ in the 2D insulating cuprates is expected to be
0.1$\sim$0.2,\cite{Tohyama,Eskes,Annett,Morr} the ground state
remains to be the N\'eel state.  Although the N\'eel state is stable,
the excited states obtained after kicking photoelectron out of the
system might be sensitive to the magnetic frustration.  As far as we
know there has been no study to examine the effect of $J'$ on the
single-hole spectral function.

In this paper, we examine the effect of the magnetic frustration
$J'$ on the single-hole spectral function at zero and finite
temperatures in the $t$-$t'$-$t''$-$J$ model.  At zero temperature,
the exact diagonalization (ED) and self-consistent Born approximation
(SCBA) methods are used.  The frustration is found to suppress the
QP weight at small momentum.  The suppression of the QP peak gives
rise to a broad spectrum at the momentum, whereas the QP peak at
($\pi$/2,$\pi$/2) remains sharp.  This is consistent with the ARPES
data on Sr$_2$CuO$_2$Cl$_2$ that the ($\pi$/2,0) spectrum is broader
than the ($\pi$/2,$\pi$/2) one.  We also calculate the single-hole
spectral function at finite temperatures in the $t$-$t'$-$t''$-$J$
model with $J'$ by using the ED method.  It is shown that the
lineshapes between the ($\pi$/2,0) and ($\pi$/2,$\pi$/2) spectra
behave differently with increasing temperature.  This difference
is also consistent with experimental data by Kim and Shen.\cite{Kim2}
These findings indicate the importance of the magnetic frustration
on the electronic states of the 2D insulating cuprates.

\section{Method of calculations}

The single-hole spectral function at temperature $T$ is given by
\begin{eqnarray}
A({\bf k},\omega)&=&
{1\over \cal{Z}}\sum\limits_{m,n} \left|\left< \Psi_n^{N-1} \left|
c_{{\bf k}\sigma}
\right| \Psi_m^N \right>\right|^2 \nonumber\\
& & {} \times\delta(\omega- E_m^N+E_n^{N-1})e^{-\beta E_m^N},
\label{A}
\end{eqnarray}
where $\Psi_m^N$ is the wave function of the $m$-th eigenstates
with energy $E_m^N$ in the $N$-electron system.
${\cal{Z}}=\sum_m e^{-\beta E_m^N}$ is the partition function with
$\beta=1/k_{\rm B}T$.  The Boltzmann factor $k_{\rm B}$ is taken
to be 1, hereafter.  The initial states $\Psi_m^N$ are the
eigenstates of the $J$-$J'$ Heisenberg model.  At $T$=0, the initial
state is restricted to the ground state of the $J$-$J'$ model.
In the evaluation of Eq.~(\ref{A}) at finite temperatures, we use
a 4$\times$4 lattice with periodic boundary condition.  After
diagonalizing the $J$-$J'$ Hamiltonian by the Householder method,
we make use of the Lanczos procedure for each initial eigenstate
in the Hamiltonian to obtain $A({\bf k},\omega)$.

At $T$=0, we also use the self-consistent Born approximation (SCBA)
method to calculate $A({\bf k},\omega)$.\cite{Schmitt,Kane,Martinez}
In the SCBA method, the hole Green's function
$G\left({\bf k},\omega\right)$ and the self-energy
$\Sigma\left({\bf k},\omega\right)$ satisfy a set of self-consistent
equation for a 2D square lattice with $N$ sites:
\begin{eqnarray}
G\left({\bf k},\omega\right) &=& {1\over\omega-\Sigma\left({\bf k},
\omega\right) -\varepsilon_{\bf k}+i\eta}\;, \\
\label{G}
\Sigma\left({\bf k}, \omega\right) &=& \sum\limits_{{\bf q}}
F\left({\bf k},{\bf q}\right) G\left({\bf k}-{\bf q},
\omega-\omega_{\bf q} \right) \;, \\
\label{SelfEnergy}
F\left({\bf k}, {\bf q} \right) &=& {32t^2\over N}\left|
\gamma_{{\bf k} -{\bf q}}u_{\bf q} +\gamma_{\bf k} v_{\bf q}
\right|^2 \;,
\label{Fkq}
\end{eqnarray}
where
\begin{eqnarray}
\varepsilon_{\bf k} &=& 4t'\cos k_x \cos k_y +2t'' \left(\cos 2k_x
+ \cos 2k_y\right)\;, \\
\label{Ek}
\omega_{\bf q}&=&2J\sqrt{{\jmath_{\bf q}}^2-{\gamma_{\bf q}}^2} \; , \\
\label{Omegaq}
\jmath_{\bf q}&=&1-{J'\over J}\left(1-{\rm cos}q_x{\rm cos}q_y\right) \; , \\
\label{jeff}
\gamma_{\bf k}&=&{1 \over2}\left({\rm cos}k_x+{\rm cos}k_y\right) \; , \\
\label{gammaq}
u_{\bf q}&=&{1\over \sqrt{2}}\left(\sqrt{{{\jmath_{\bf q}}^2
\over{{\jmath_{\bf q}}^2-{\gamma_{\bf q}}^2}}} +1\right)^{1 \over2} \; , \\
\label{uq}
v_{\bf q}&=&-{\rm sign}\left(\gamma_{\bf q}\right){1\over \sqrt{2}}
\left(\sqrt{{{\jmath_{\bf q}}^2 \over{{\jmath_{\bf q}}^2-
{\gamma_{\bf q}}^2}}} -1\right)^{1 \over2} \; ,
\label{vq}
\end{eqnarray}
and 1$\gg$$\eta$$>$0.
The AF long-range order is assumed in the derivation of these equations.
In the next section, we will compare the SCBA results with ED ones and
see the validity of the assumption.  $A\left({\bf k},\omega\right)$ is
assumed to be equal to the imaginary part of the hole Green's function
Eq.~(7):
\begin{equation}
A\left({\bf k},\omega\right)=-{1\over \pi}{\rm Im} G
\left(k,\omega\right)\;.
\end{equation}

\begin{figure}[b]
\epsfxsize=8.5cm
\centerline{\epsffile{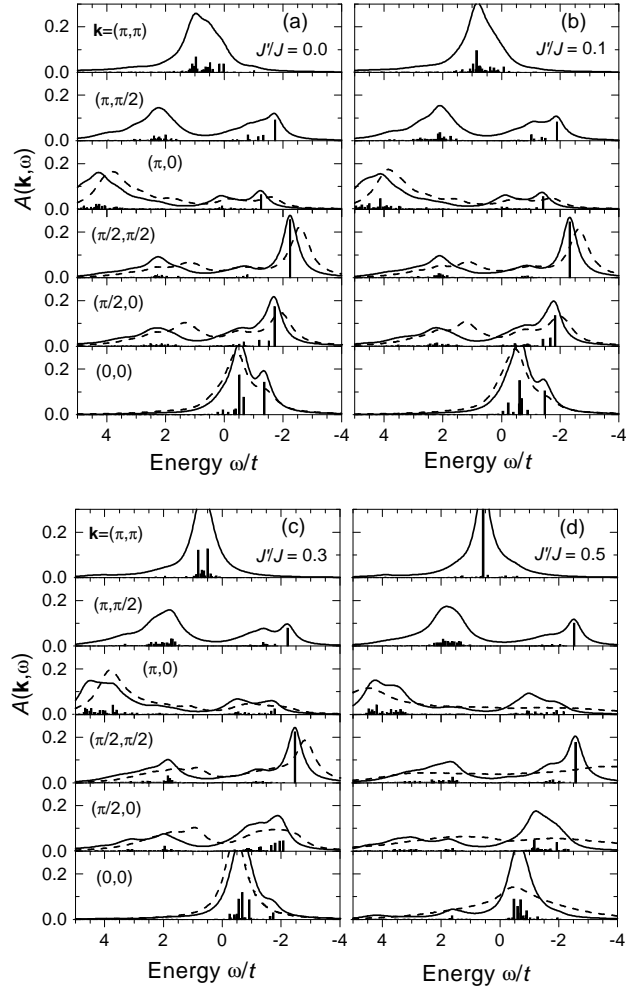}}
\vspace{4mm}
\caption{Single-hole spectral function $A({\bf k},\omega)$ of the
$t$-$t'$-$t''$-$J$ model with various values of $J'$ at $T$=0.
$t'$=$-$0.34$t$, $t''$=0.23$t$ and $J$=0.4$t$.  $J'$/$J$ is 0 (a),
0.1 (b), 0.3 (c), and 0.5 (d).  Calculations are performed on the
$4\times4$ cluster by the ED method (solid line) and the SCBA method
(dashed line).  The SCBA data at ($\pi$,$\pi$) and ($\pi$,$\pi$/2)
are excluded from the panels, since they are equivalent to those at
(0,0) and ($\pi$/2,0), respectively.  The solid lines are obtained
by introducing a Lorentzian broadening of 0.4$t$ into
$\delta$-functions (vertical bars) in the ED results.}
\label{Fig1}
\end{figure}

\section{Results and discussions}

As mentioned in Sec.~I, the nature of the ground state in the
$J$-$J'$ Heisenberg model changes from the N\'eel to collinear
states at $J'/J\sim$0.6.\cite{Mila,Miyazawa}  By using the ED method,
we found that the single-hole spectral function also shows different
behavior between the two states.\cite{Shibata}  However, since we are
interested in hole dynamics in the AF spin system with realistic
values of $J'$/$J$, we focus on the region $J'$/$J$$<$0.6.

\subsection{Zero Temperature}

Figure~\ref{Fig1} shows $A({\bf k},\omega)$ of the $t$-$t'$-$t''$-$J$ model
with various values of $J'$ at $T$=0.  Parameter values are set to
be $J$=0.4$t$, $t'$=$-$0.34$t$ and $t''$=0.23$t$ as used in Ref.~2.
Solid lines and vertical bars are the ED results on a 4$\times$4
cluster.  In Fig.~1(a) ($J'$/$J$=0), the height of the QP peak
($\omega$/$t$$\sim$$-$1.2) at ($\pi$,0) is small as compared with
those at ($\pi/2,\pi/2$) and ($\pi/2$,0) ($\omega$/$t$$\sim$$-$2.2
and $-$1.7, respectively).\cite{Kim}  Since $t'$ and $t''$ terms
cause the shift of the QP peak at ($\pi$,0) to higher binding energy,
the position of the QP peak enters into the region where the incoherent
states exist.  As a result, the QP weight is strongly suppressed.

With increasing $J'$ the ED results show the decrease of the QP
weight for all the momenta.  In particular, the remarkable decrease
is found at (0,0) and ($\pi$/2,0).  The spectrum near the QP energy
at ($\pi$/2,0) becomes broad as clearly seen in Figs.~1(c) and (d).
Such a broad spectrum has been observed at ($\pi$/2,0) in ARPES
data.\cite{Wells,Kim}  Since the QP peak without $J'$ is as sharp as
that at ($\pi$/2,$\pi$/2) (see Fig.~1(a)), the observed broad spectrum
is considered to be a manifestation of the effect of frustration due to $J'$.

In order to clarify the effect of $J'$ on the spectral function, we
also use the SCBA method for the calculation of $A({\bf k},\omega)$.
In Fig.~1, the SCBA results are compared with the ED ones to see the
validity of the approximation.  The SCBA data (dashed lines) are
plotted for momenta inside the magnetic first Brillouin zone.  The
spectra obtained by the SCBA method are in good agreement with those
by the ED one, except for $J'$/$J$=0.5 where the N\'eel state is
expected to be less stable in the ground state.
Encouraged by the good agreement between the ED and SCBA results at
$J'$/$J$$\leq$0.3, we extend the size of cluster for the SCBA method
from 4$\times$4 to 16$\times$16.  Figure~2 shows the SCBA results at
various momenta for the 16$\times$16 cluster.  Shown in the figure are
$A({\bf k},\omega)$, the imaginary part of the self-energy
${\rm Im} \Sigma\left({\bf k},\omega\right)$ and the QP weight
$Z\left({\bf k},\varepsilon_{\rm QP}\right)$ defined as
\begin{equation}
Z\left({\bf k},\varepsilon_{\rm QP}\right)={\left(1-
{{\partial{\rm Re}\Sigma\left({\bf k},\omega\right)\over
\partial\omega}\Bigl|}_{\omega=\varepsilon_{\rm QP}}\right)}^{-1}\;,
\end{equation}
where $\varepsilon_{\rm QP}$ is the QP energy at ${\bf k}$.

\begin{figure}
\epsfxsize=8.5cm
\centerline{\epsffile{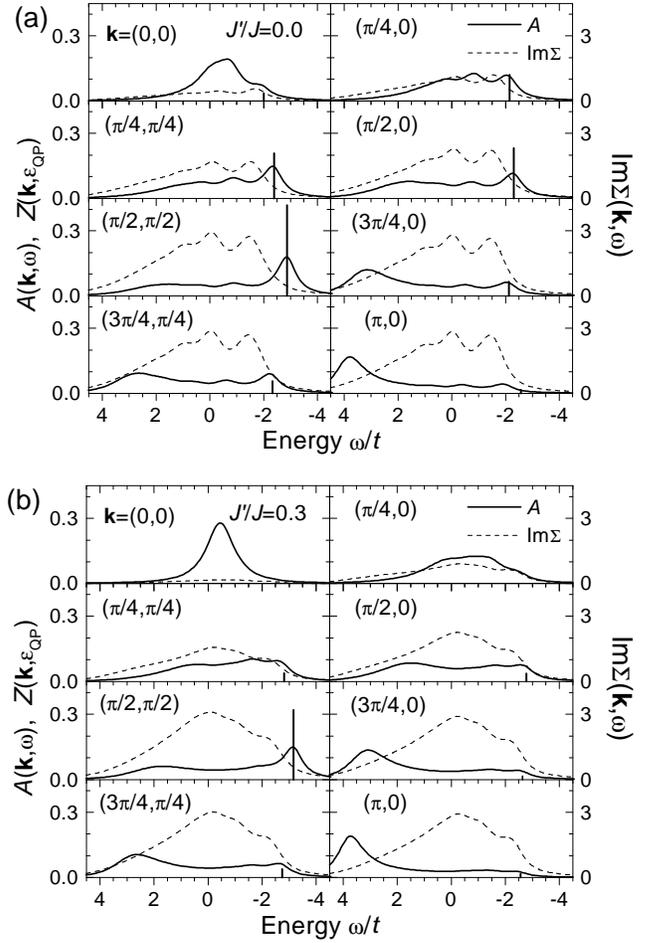}}
\vspace{4mm}
\caption{Single-hole spectral function $A({\bf k},\omega)$
(solid line), the imaginary part of the self-energy
${\rm Im} \Sigma\left({\bf k},\omega\right)$ (dashed line) and
the QP weight $Z\left({\bf k},\varepsilon_{\rm QP} \right)$
(vertical line) of the $t$-$t'$-$t''$-$J$ model with $J'$ at $T$=0
obtained by the SCBA method on a 16$\times$16 cluster.
$t'$=-0.34$t$, $t''$=0.23$t$ and $J$=0.4$t$.  $J'$/$J$ is 0 (a)
and 0.3 (b).  $\eta$/$t$=0.4 for $A\left({\bf k},\omega\right)$ and
${\rm Im}\Sigma\left({\bf k},\omega\right)$, and $\eta$/$t$=0.005
for $Z\left({\bf k},\varepsilon_{\rm QP}\right)$.  In (b), $Z$ at
(0,0) and ($\pi$/4,0) is zero.}
\label{Fig2}
\end{figure}

With increasing $J'$/$J$ from 0 (Fig.~2(a)) to 0.3 (Fig.~2(b)),
the spectral intensity around the QP energy becomes broader when
the momentum {\bf k} is close to (0,0).  Since the life time of
the QP expressed by ${\rm Im} \Sigma\left({\bf k},\omega\right)$
shows little change, the broadness comes from the decrease of the
QP weight $Z$.  For {\bf k}=(0,0) and ($\pi$/4,0), $Z$ is zero at
$J'$/$J$=0.3.  On the other hand, for large momenta such as
(3$\pi$/4,$\pi$/4) and ($\pi$,0), the change of $Z$ is very small.
In the intermediate values of {\bf k}, the ratio of $Z$ for
$J'$/$J$=0.3 to that for $J'$/$J$=0 decreases as ${\bf k}$
approaches (0,0) point.  The maximum value of the ratio appears
at ($\pi$/2,$\pi$/2).  Similar {\bf k} dependent behaviors of $Z$
are obtained when the value of $J$ in the $t$-$J$ model is
reduced.\cite{Martinez}  The similarity between the effects due to
$J$ in the $t$-$J$ model and $J'$ in our model is also seen in the
dependence of the QP band width $W$, i.e. with increasing $J'$,
$W$ decreases monotonically, which resembles a well-known behavior
of $W$$\propto$$J$ in the $t$-$J$ model.\cite{Dagotto2}  Therefore,
we can say that the changes of $A({\bf k},\omega)$ due to the
increase of $J'$ are roughly similar to those due to the decrease of
$J$ in the $t$-$J$ model.  This is consistent with an intuitive
picture that $J'$ frustrate the N\'eel state as if the net value of
$J$ is reduced.

\begin{figure}
\vspace{2mm}
\epsfxsize=8.5cm
\centerline{\epsffile{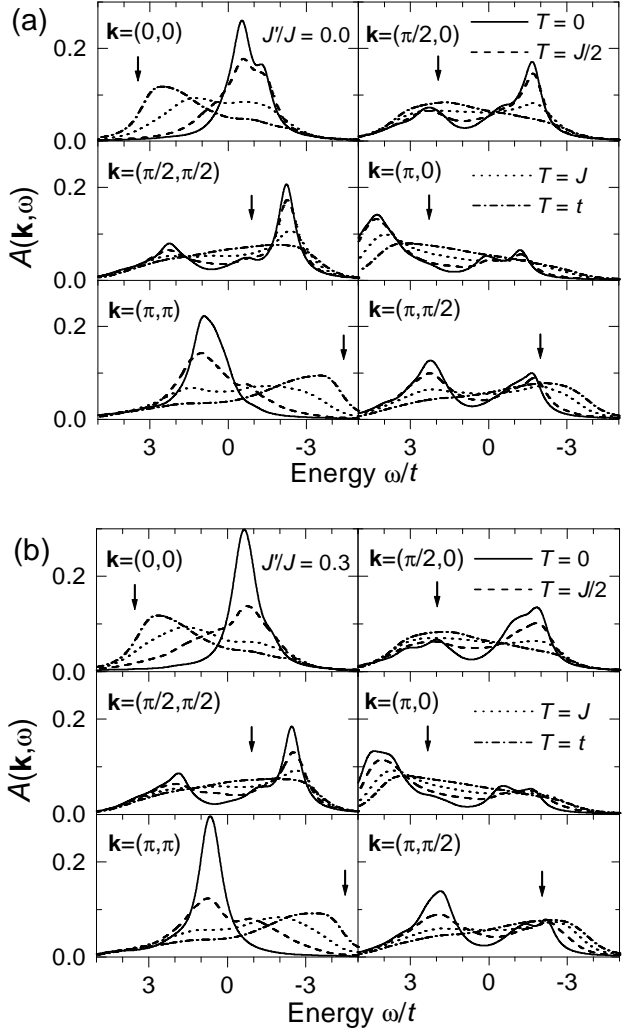}}
\vspace{4mm}
\caption{Single-hole spectral function $A({\bf k},\omega)$ of the
$t$-$t'$-$t''$-$J$ model with $J'$ at various temperatures.
$t'$=$-$0.34$t$, $t''$=0.23$t$ and $J$=0.4$t$.  $J'$/$J$ is 0 (a)
and 0.3 (b).  The values of temperature are $T$=0 (solid line),
$J$/2 (dashed line), $J$ (dotted line) and $t$ (dot-dashed line).
Calculations are performed on the $4\times4$ cluster by the ED
method.  A Lorentzian broadening of 0.4$t$ is introduced into the
$\delta$-functions.}
\label{Fig3}
\end{figure}

\subsection{Finite Temperatures}

We next present the temperature dependence of $A({\bf k},\omega)$
on a 4$\times$4 lattice.  Since the energy separation between
the ground and first excited states in the $J$-$J'$ Heisenberg model
on the lattice is roughly $J$/2, reliable results are obtained only
in the region of $T\agt J$/2.  Although the temperature one can
treat is rather high, the effect of the change of spin correlation
length $\xi$ on the spectrum may be seen in the calculation
because the correlation length $\xi$ for the temperatures above
$T$$\sim$$J$/2 is less than the maximum length of the 4$\times$4
lattice ($\sim2\sqrt{2}a$).\cite{Ding}

Figure~3 shows $A({\bf k},\omega)$ of the $t$-$t'$-$t''$-$J$ model
with $J'$/$J$=0 and 0.3 at various temperatures obtained by using
the ED method.  The QP spectral intensity at ($\pi$/2,$\pi$/2)
decreases with increasing temperature.  This is not due to the
decrease of the QP weight but due to the decrease of the life
time.\cite{Jaklic}  We find that, when the temperature changes
from $T$=0 to $J$/2, the intensity in Fig.~3(b) ($J'$/$J$=0.3)
is largely suppressed as compared with that in Fig.~3(a).
This simply comes from the fact that the frustration $J'$ disturbs
the AF spin background which is responsible for the formation of QP
and makes the background sensitive to a slight change in
temperature.  Such a sensitive behavior of the QP peak may be one
of the origins of the observed strong temperature dependence of
ARPES data.\cite{Kim2}

In Fig.~3(b), we find that the dependence of the spectra near the
QP energy on temperature at ($\pi$/2,$\pi$/2) is remarkably
different from that at ($\pi$/2,0). At ($\pi$/2,$\pi$/2) the
spectral weight around  $\omega$$\sim$$-$$t$ slightly increases
with increasing temperature in spite of the reduction of the QP
intensity.  On the contrary, the weight in the same energy region
at ($\pi/2$,0) is uniformly suppressed with increasing temperature.
Such a different behavior has been observed in ARPES experiment on
Sr$_2$CuO$_2$Cl$_2$.\cite{Kim2}.  Since the behavior can not be
clearly seen without $J'$ (see Fig.~3(a)), the agreement between
theory and experiment indicates the importance of $J'$ on the
electronic states of insulating cuprates.

With increasing temperature from $T$=0 to $J$, the QP spectral
intensity in Figs.~3(a) and (b) rapidly decreases accompanied
by a slight shift to lower-energy side.\cite{Igarashi,Brink,Prelovsek}
At $T$=$J$, the QP peaks almost vanish because the correlation
length $\xi$ is less than one-lattice spacing.\cite{Ding}  Similar
suppression of the weight as the temperature increases from $T$=0
to $J$ is seen in the other peaks at $\omega\sim$0, $t$, 2$t$ and
4$t$ for (0,0), ($\pi$,$\pi$), ($\pi$,$\pi$/2) and ($\pi$,0),
respectively.  The weight at these peaks is thus sensitive to the
nature of the spin background.  At $T$=$t$, the spectrum has a
broad maximum at a certain energy in each momentum.  This behavior
is independent of the value of $J'$.  In such a high temperature
region, the hole can propagate without being disturbed by the spin
background.\cite{Brinkman}  This produces a dispersive structure
whose energy is determined by a tight-binding band due to the free
hole propagation.  The energy band is given by
\begin{eqnarray}
E_{\rm free}\left({\bf k}\right)&=&2t\left(\cos k_x+\cos k_y\right)
+4t'\cos k_x \cos k_y \\ \nonumber
&& +2t''\left(\cos 2k_x+ \cos 2k_y\right)\;.
\label{Enon}
\end{eqnarray}
In Fig.~3, the energy obtained from the band is indicated by arrow
at each {\bf k} point.  With further increasing temperature,
the maximum of the spectrum is expected to shift to the position
indicated by the arrow.

\section{Conclusions}

We have examined the effect of the magnetic frustration $J'$ on
the single-hole spectral function in the $t$-$t'$-$t''$-$J$ model,
by using the ED and SCBA methods.  At zero temperature, the
frustration suppresses the QP weight and makes the spectrum broad
for small momentum, whereas the peak at ($\pi$/2,$\pi$/2) remains
sharp.  Our results explain the experimental data that the
($\pi$/2,0) spectrum in Sr$_2$CuO$_2$Cl$_2$ is broader than the
($\pi$/2,$\pi$/2) one.  We have also evaluated the temperature
dependence of the single-hole spectral function in the
$t$-$t'$-$t''$-$J$ model with $J'$ by using the ED method.
The lineshape between ($\pi$/2,0) and ($\pi$/2,$\pi$/2) spectra
behaves differently with increasing temperature.  This difference
is also consistent with recent experimental data.  These findings
indicate the importance of the magnetic frustration on the
electronic states of the insulating cuprates.

\acknowledgements
We would like to thank C. Kim and Z.-X. Shen for sharing with their
experimental data prior to publication and for stimulating discussions.
We also thank P. Prelov\v{s}ek for informing us Ref.~26.
This work was supported by CREST and NEDO.
The numerical calculations were performed in the Supercomputer Center,
Institute for Solid State Physics, University of Tokyo, and the
supercomputing facilities in Institute for Materials Research,
Tohoku University.


\begin{references}

\bibitem{Wells} B. O. Wells, Z.-X. Shen, A. Matsuura, D. M. King,
M. A. Kastner, M. Greven, and R. J. Birgeneau, Phys. Rev. Lett. {\bf 74},
964 (1995).
\bibitem{Kim} C. Kim, P. J. White, Z.-X. Shen, T. Tohyama, Y. Shibata,
S. Maekawa, B. O. Wells, Y. J. Kim, R. J. Birgeneau, and M. A. Kastner,
Phys. Rev. Lett. {\bf 80},4245 (1998).
\bibitem{Nazarenko} A. Nazarenko, K. J. E. Vos, S. Haas, E. Dagotto, and
R. Gooding, Phys. Rev. B {\bf 51}, 8676 (1995).
\bibitem{Kyung} B. Kyung and R. A. Ferrell, Phys. Rev. B {\bf 54}, 10125
(1996).
\bibitem{Xiang} T. Xiang and J. M. Wheatley, Phys. Rev. B {\bf 54},
R12653 (1996).
\bibitem{Belinicher} V. I. Belinicher, A. L. Chernyshev, and V. A. Shubin,
Phys. Rev. B {\bf 54} 14914 (1996).
\bibitem{Lee} T. K. Lee and C. T. Shih, Phys. Rev. B {\bf 55}, 5983 (1997).
\bibitem{Eder} R. Eder, Y Ohta, and G. A. Sawatzky, Phys. Rev. B {\bf 55},
R3414 (1997).
\bibitem{Leung1} P. W. Leung, B. O. Wells, and R. J. Gooding, Phys. Rev. B
{\bf 56}, 6320 (1997).
\bibitem{Lema} F. Lema and A. A. Aligia, Phys. Rev. B {\bf 55}, 14092
(1997).
\bibitem{Duffy} D. Duffy, A. Nazarenko, S. Haas, A. Moreo, J. Riera, and
E. Dagotto, Phys. Rev. B {\bf 56}, 5597 (1997).
\bibitem{Anderson}P. W. Anderson, Science {\bf 235}, 1196 (1987).
\bibitem{Mila} F. Mila, D. Poilblanc, C. Bruder, Phys. Rev. B {\bf 43}, 7891
(1991).
\bibitem{Miyazawa}S. Miyazawa and S. Homma, Phys. Lett. A {\bf 193}, 370
(1994).
\bibitem{Tohyama}T. Tohyama and S. Maekawa, J. Phys. Soc. Jpn. {\bf
59}, 1760 (1990).
\bibitem{Eskes}H. Eskes, L. F. Feiner and G. A. Sawatzky, Physica C
{\bf 160}, 424 (1989)
\bibitem{Annett}J. F. Annett, R. M. Martin, A. K. McMahan and S. Satpathy,
Phys. Rev. B {\bf 40}, 2620 (1989).
\bibitem{Morr}D. K. Morr, Phys. Rev. B {\bf 58}, R587(1998).
\bibitem{Kim2}C. Kim and Z.-X. Shen, private communication.
\bibitem{Schmitt}S. Schmitt-Rink, C. M. Varma, and A. E. Ruckenstein,
Phys. Rev. Lett. {\bf 60}, 2793 (1988).
\bibitem{Kane}C. L. Kane, P. A. Lee and N. Read, Phys. Rev. B
{\bf 39}, 6880 (1989).
\bibitem{Martinez}G. Martinez and P. Horsh, Phys. Rev. B {\bf 44},
317 (1991).
\bibitem{Shibata}Y. Shibata, T. Tohyama and S. Maekawa, unpublished.
\bibitem{Dagotto2}E. Dagotto, Rev. Mod. Phys. {\bf 66}, 763 (1994);
and references therein.
\bibitem{Ding}H.-Q. Ding and M. S. Makivi\'c, Phys. Rev. Lett., {\bf 64},
1449 (1990).
\bibitem{Jaklic} J. Jakli\v{c}, thesis, University of Ljubljana (1996).
\bibitem{Igarashi} J. Igarashi and P. Fulde, Phys. Rev. B {\bf 48},
998 (1993).
\bibitem{Brink} J. van den Brink and O. P. Sushkov, Phys. Rev. B
{\bf 57}, 3518 (1998).
\bibitem{Prelovsek}J. Jakli\v{c} and P. Prelov\v{s}ek, cond-mat/9803331
(to be publised in Adv. Phys.).
\bibitem{Brinkman}W. F. Brinkman and T. M. Rice, Phys. Rev. B {\bf 2},
1324 (1970).
\end{references}
\end{document}